\documentstyle[11pt]{article}
\newtheorem{thm}{Theorem}[section]
\newtheorem{cor}[thm]{Corollary}

\newtheorem{prop}[thm]{Proposition}
\begin{document}

\title{\bf  {Harmonic and Wave Maps Coupled with Einstein's
gravitation}}
\author{{R. Schimming$^{1}$\thanks{%
E-mail: rschimming@t-online.de}   and Ragab M. Gad$^{2,3}$ \thanks{%
E-mail: ragab2gad@hotmail.com} }\\
\newline
{\it $^1$ Universit\"{a}t Graeifswald, Institut f\"{u}r Mathematik
und informatik,}\\
 {\it  D-17487 Greifswald, Germany}
 \\
{\it $^2$
 Mathematics Department, Faculty of Science, King
Abdulaziz University,}\\
 {\it  21589 Jeddah, KSA}
 \\
{\it $^3$ Mathematics Department, Faculty of Science, Minia University,}\\
 {\it   61915 El-Minia,  Egypt}
}

\date{\small{}}

\maketitle

\begin{abstract}
In this paper we discuss the coupled dynamics, following from a
suitable Lagrangian, of a harmonic or wave map $\phi$ and Einstein's
gravitation described by a metric $g$. The main results concern
energy conditions for wave maps, harmonic maps from warped product
manifolds, and wave maps from wave-like Lorentzian manifolds.
\end{abstract}
{\bf{Keywords}}: Harmonic map; Wave map; Energy conditions.
%%% ----------------------------------------------------------------------
%\maketitle
%%% ----------------------------------------------------------------------
\setcounter{equation}{0}
\section{Introduction}
Scalar fields on a space or space-time manifold $(X, g)$, which satisfy
a linear or nonlinear field equation, attract enduring attention; cf, eg., \cite{1,2,3}.
The {\it{Dirichlet Lagrangian}} or {\it{energy density}}
\begin{equation}
 e=\frac{1}{2} |d\phi |^2 = \frac{1}{2} g^{ab}(\partial_a\phi )(\partial_b \phi )
\end{equation}
of a one-component scalar field $\phi = \phi (x)$ leads to a linear field equation of Laplace or D'Alembert type. Here we denote
$$
g = g_{ab}dx^a dx^b, \quad (g^{ab}):= (g_{ab})^{-1}, \quad \partial_a : = \frac{\partial}{\partial x^a}.
$$
The Dirichlet Lagrangian $e$ admits a natural generalization to a multi-component scalar field $\phi = \phi (x)$ if the range of $\phi$ is suitably geometrized, namely if $im \, \phi$ lies in a Riemannian manifold $(Y, h)$. That means, $\phi$ becomes a map $X \rightarrow Y$, $x \mapsto y$ between  a source manifold $(X, g)$ and a target manifold $(Y, h)$. The choice of the Lagrangian
\begin{equation}\label{1.1-}
e = e[\phi ]= e[\phi , g]= \frac{1}{2} |d\phi |^2,
\end{equation}
where now
$$
 |d\phi|^2:= tr(\phi^*h) \equiv g^{ab} (\partial_a\phi^i)(\partial_b\phi^j)h_{ij}(\phi^k),
$$
leads to the generally semi-linear field equation with Laplace-like or D'Alembert-like principal part
\begin{equation}\label{1}
 tr (\nabla d\phi )^i \equiv g^{ab}\nabla_a \partial_b \phi^i =0.
\end{equation}
Here we denote $h= h_{ij}dy^idy^j$
and  the special covariant derivative $\nabla$ is built from $g$, $h$, $\phi$ as will be explained in Section 2.
\par
A map $\phi : X \longrightarrow Y$ between properly Riemannian manifolds $(X, g)$, $(Y, h)$ which satisfies (\ref{1}) is called a {\it{harmonic map}} \cite{4,5}. A map $\phi : X \longrightarrow Y$ from a Lorentzian manifold $(X, g)$ to a properly Riemannian manifold $(Y, h)$ which satisfies (\ref{1}) is called a {\it{wave map}} \cite{3,6,7,8,9}.
\par
In this paper, we take $(Y, h)$ as fixed background and consider $g$ and $\phi$ as dynamical objects. The dynamics shall follow from the Lagrangian
\begin{equation}\label{2}
 L = \kappa R - e,
\end{equation}
where $R$ denotes the scalar curvature of $g$ and $\kappa \neq 0$ is a coupling constant.\\
Variation with respect to $g$ yields, for $dim X \geq 3$, the {\it{Einstein equation}} in the form
\begin{equation}\label{3}
 \kappa Ric = \phi^* h,
\end{equation}
in components
\begin{equation}\label{4}
 \kappa R_{ab} = (\partial_a \phi^i)(\partial_b\phi^j)h_{ij}.
\end{equation}
Variation  with respect to $\phi$ yields (\ref{1}).
\par
A Lorentzian metric $g$ can be interpreted as {\it{gravitation}}. According to the Kaluza-Klein principle, the space-time manifold $(X, g)$ may have a higher dimension.
\par
A positive definite metric $g$ on $X$ can be given a physical interpretation through a Lorentzian metric constructed from it  as follows. Consider the product manifold $\tilde{X} := \Re \times X$ with points $(t, x)$ and equip it with the Lorentzian metric $\tilde{g} := dt^2 -g$. Extend $\phi : X \longrightarrow Y$ to $\tilde{\phi} : \tilde{X} \longrightarrow Y$ by setting $\tilde{\phi}(t, x) : = \phi(x)$, that means $\tilde{\phi} := \phi \circ pr_2$, where $pr_2$ is the projection $\tilde{X} \longrightarrow X$ to the second factor. The field equations (\ref{3}), (\ref{1}) for $(\tilde{X}, \tilde{g})$, $\tilde{\phi}$ reduce to (\ref{3}), (\ref{1}) for $(X, g)$, $\phi$.\\

Let us sketch our main results.
\begin{itemize}
\item  The energy-momentum tensor $T$ of a wave map obeys several energy conditions.
In particular, if $v$ is a causal (i.e. non-spacelike) vector field then
the momentum one-form  $I :=T(\cdot,v)$  is causal again.
\item  The Einstein equation (1.5) implies that the conditions $Ric(v,v) = 0$,
$Ric(\cdot,v) = 0$, $\phi_*v = 0$ for a vector field $v$ are equivalent to each other.
Moreover, then $v$ is a Ricci collineation, i.e. $\pounds_v Ric = 0$. Some conclusions
are drawn from this latter fact.
\item A submersive map $\phi :X \rightarrow Y$ between a pure manifold $X$ and a Riemannian
manifold $(Y,h)$ can locally be made to a harmonic or wave map by a
suitable choice of a metric $g$ on $X$.
\item  We study the case of warped product $X = {^\prime}X \times {^{\prime\prime}}X$,  $g = {^\prime}g \oplus w^2\, {^{\prime\prime}}g$.
Several propositions are proved by means of the argument that the integral
of a Laplace expression over a closed manifold $^{\prime}X$ vanishes.
\item  We study radiation conditions for a Lorentzian manifold $(X,g)$. The Einstein
equation (1.5) leads from one condition to a stronger condition.
\end{itemize}

\setcounter{equation}{0}
 \section{Preliminaries}
We consider dimensions
$$
m: =dim X \geq 3, \quad n : = dim Y \geq 1
$$
and adopt the following index convention: indices $a, b, c, ...$ label the components of geometric objects on $X$; indices $i, j, k, ...$ label the components of geometric objects on $Y$.\\
Tensor fields on $Y$ become multi-component scalar fields on $X$ by insertion of $y = \phi(x)$, where $x\in X$, $y\in Y$. For covariant tensor fields on $Y$ there is also the conventional pull-back map $\phi^*$. For instance, the metric $h=h_{ij}dy^idy^j$
yields $h\circ\phi$ with components  $h_{ij}(\phi^k) = h_{ij}(\phi^k(x^a))$ on $X$ and also the pull-back $\phi^*h$ on $X$ with components
$(\partial_a\phi^i)(\partial_b\phi^j)(h_{ij}(\phi^k)$. The object $\phi^*h$ is called {\it{first fundamental form}} of $\phi : X\longrightarrow Y$.\\
We will occasionally  write
$$
\phi^*h = h(d\phi, d\phi )
$$
and also use the bilinear symmetric form $h(\cdot , \cdot )$ with tensorial entries.\\
Some natural covariant derivative $\nabla$ with components $\nabla_a$ is built from $g$, $h$, $\phi$ according to the following rules
\begin{enumerate}
 \item $\nabla$ applied to tensor fields on $X$ equals the Levi-Civita derivative $^g\nabla$ to $g$. For instance,
$$
\nabla_a v^c := \partial_av^c + {^g}\Gamma_{ab}^c v^b,
$$
where $v = v^a\partial_a$ is a vector field on $X$ and $^g\Gamma^c_{ab}$ are the Christoffel symbols to $(g_{ab})$.
\item $\nabla$ applied to tensor fields on $Y$ equals some pull-back of the Levi-Civita derivative $^h\nabla$ to $h$. For instance,
$$
\nabla_a w^k := (\partial_a\phi^i)(\partial_iw^k + {^h}\Gamma_{ij}^k w^j),
$$
where $w=w^i\partial_i$ is a vector field on $Y$ and $^h\Gamma_{ij}^k$ are the Christoffel symbols to $(h_{ij})$. Here $\nabla_aw^k$ is understood to depend on $y^l = \phi^l(x^c)$.
\item There are natural product rules for mixed quantities, the components of which carry both indices $a, b, ...$ and  $i, j, ...$. For instance, the {\it{differential}} $d\phi$ of $\phi : X \longrightarrow Y$ with components $\partial_a\phi^i$ is a mixed tensor field. The covariant derivative $\nabla d\phi$ of $d\phi$ is called {\it{second fundamental form}} of $\phi$. It has the components
$$
\nabla_a\partial_b\phi^k=\partial_a\partial_b\phi^k - {^g}\Gamma_{ab}^c \partial_c\phi^k + {^k}\Gamma^h_{ij}(\partial_a\phi^i)(\partial_b\phi^j)
$$
and the symmetry property
$$
\nabla_a\partial_b\phi^k= \nabla_b\partial_a\phi^k.
$$

More on calculus for maps between (pseudo-) Riemannian manifolds can be
found in the literature, e.g. \cite{4,5}.

\end{enumerate}

\setcounter{equation}{0}
 \section{The field equations}
The field theory for $g$ and $\phi$ considered here is based on the Lagrangian
\begin{equation}\label{6}
 L=\kappa R - e,
\end{equation}
which is the sum of the {\it{gravitational Lagrangian}} $\kappa R$ and the {\it{matter Lagrangian}} $e$. Here $R= R[g]$ denotes the scalar curvature of $g$, $\kappa \neq 0$ is a coupling constant and $e=e[\phi ] = e[g, \phi ]$ is given by (\ref{1.1-}).\\
The idea to couple a harmonic map, formerly also called {\it{sigma model}}, with
Einstein's gravitation appeared in \cite{10,11,12} and other early papers.\\
The following is well-known \cite{7,8,9,10,11,12,13,14,15,16,17,18}. We
abbreviate $\det g :=\det (gab)$.
\begin{prop}\label{pro1}
 Variation of $L\mid\det g\mid^{\frac{1}{2}}$  with respect to $g$ yields the Einstein equation in the form
\begin{equation}\label{7}
\kappa (Ric - \frac{1}{2}Rg) = \phi^*h - eg.
\end{equation}
If $m = dim X \geq 3$ then this is equivalent to
\begin{equation}\label{7-}
\kappa Ric = \phi^* h.
\end{equation}
Variation of $L\mid\det g\mid^{\frac{1}{2}}$  with respect to $\phi$ yields
\begin{equation}\label{7--}
tr (\nabla d\phi )=0,
\end{equation}
where $\nabla d\phi$ is the second fundamental form of $\phi$ and the trace $tr$ refers to the metric $g$.
\end{prop}
The right-hand side of (\ref{7})
\begin{equation}\label{8}
 T:= \phi^*h - eg
\end{equation}
is called {\it{energy-momentum tensor of $\phi$}}. There holds $e=\frac{1}{2} tr (\phi^*h)$ and (\ref{8}) is equivalent to
$$
T - (m-2)^{-1}(tr T)=\phi^*h.
$$
\begin{prop}\label{pro2}
From the field equation (\ref{7--}) for $\phi$ there follows that $T$ is divergence-free:
$$
\nabla^bT_{ab}=0.
$$
\end{prop}
The {\it{proof}} follows from the identity
$$
\nabla^bT_{ab}=h_{ij}(\partial_a\phi^i)(tr \nabla d\phi)^j.
$$

\setcounter{equation}{0}
 \section{Energy conditions for a wave map}
Let now the metric $g$ have Lorentzian signature $(+-...-)$.\\
%\begin{defn}\label{defin1}
{\bf{Definition 4.1}} \,\,
 Let a symmetric two-form $T=T_{ab}dx^adx^b$ on $X$ be interpreted as an energy-momentum tensor field and let $v=v^a\partial_a$ be a unit timelike vector field on $X$, i.e., $v_av^a \equiv g_{ab}v^av^b =1$. Then $T(v, v) \equiv T_{ab}v^av^b$ is called {\it{energy density}}, $I:= T(\cdot , v)$ with components $I_a:= T_{ab}v^b$ is called {\it{momentum one-form}}, and $J := I - T(v,v)v$ with components $J_a := I_a - T(v,v)v_a$ is called {\it{proper momentum one-form}}.
\par
Physically, $v$ is interpreted as the unit tangent vector to the world line of an
observer. This observer measures the quantities $T(v,v), I, J$.\\

Every unit timelike vector field $v=v^a \partial_a$ on $X$ gives rise to a positive definite metric $g^+=g^+_{ab}dx^adx^b$ on $X$ with components $g^+_{ab} =2v_av_b - g_{ab}$. The inverse $(g^{ab}_+):=(g_{ab}^+)^{-1}$ has a representation $g^{ab}_+ = 2v^av^b -g^{ab}$.
%\end{defn}
\begin{thm}\label{theor1}
 Consider the energy-momentum tensor
\begin{equation}\label{th1}
 T= \phi^*h - eg
\end{equation}
of a map $\phi : X \longrightarrow Y$ between $(X, g)$ and $(Y, h)$. The energy density equals
\begin{equation}\label{th2}
 T(v, v) = e_+ : = e [\phi , g_+] \equiv \frac{1}{2}g^{ab}_+(\partial_a\phi^i)(\partial_b\phi^j)h_{ij}.
\end{equation}
It is a positive definite quadratic form in $d\phi$. The momentum one-form $I$ obeys the estimate
\begin{equation}\label{th3}
 e^2\leq |I|^2 \leq e_+^2,
\end{equation}
where
\begin{equation}\label{*}
|I|^2 : = I_aI^a \equiv g^{ab}I_aI_b.
\end{equation}
\end{thm}
{\bf{Proof}}:\\
 Let us abbreviate $f :=\phi^*h$ with components $f_{ab} := (\phi^*h)_{ab} = (\partial_a\phi^i)(\partial_b\phi^j)h_{ij}$.
We calculate
$$
\begin{array}{ccc}
T(v, v) & = & T_{ab}v^av^b =(f_{ab} - eg_{ab})v^av^b\\
 & = & f_{ab}v^av^b-e=\frac{1}{2}(2v^av^b -g^{ab})f_{ab}  = \frac{1}{2}g^{ab}_+f_{ab} =e_+.
\end{array}
$$
The proper momentum one-form $J$ is orthogonal to $v$, that means $J_av^a =0$. Considering that, we find that
$$
\begin{array}{ccc}
0 & \leq & g^{ab}_+J_aJ_b = - g^{ab}J_aJ_b\\
 & =& -g^{ab}J_a(I_b -e_+v_b)=-g^{ab}J_aI_b \\
 & = &-g^{ab}(I_a - e_+v_a)I_b = - I_aI^a + e_+^2.
\end{array}
$$
Thus, the right-hand side inequality of (\ref{th3}) is proved. In order to prove the left-hand side of (\ref{th3}), we start with the remark that the matrix $(f_{ab})$ is positive semi-definite. Let us consider a fixed point  $x_0\in X$ and use coordinates $x^a$ such that
\begin{equation}\label{**}
v^a = \delta_0^a, \quad g^{ab}_+ = \delta^{ab}
\end{equation}
in that very point $x_0$. Actually,  such coordinates exist. The following $2\times 2$ subdeterminants of  $(f_{ab})$ are non-negative:
$$
f_{00}f_{11} - f_{10}f_{10} \geq 0,
$$
$$
f_{00}f_{22}-f_{20}f_{20} \geq 0,
$$
$$
...
$$
$$
f_{00}f_{mm} - f_{m0}f_{m0} \geq 0.
$$
Let us sum up:
\begin{equation}\label{***}
f_{00}f_{aa} - f_{a0}f_{a0} \geq 0.
\end{equation}
Here a summation convention applies to the index $a$ and the coordinate conditions (\ref{**}) are assumed. The inequality (\ref{***}) can be can be brought into a coordinate-invariant form
$$
g^{ab}_+(f_{ab}v^cv^df_{cd}- v^cf_{ac}v^df_{bd})\geq 0.
$$
Here we insert
$$
g^{ab}_+f_{ab} = 2e_+, \quad v^cv^df_{cd}=e +e_+,
$$
$$
v^cf_{ac} = v^c (T_{ac}+eg_{ac})=I_a +ev_a,
$$
$$
g^{ab}_+(I_a +ev_a)(I_b+ev_b)=2e_+(e+e_+)+e^2-|I|^2.
$$
Taking all this together, the left-hand side inequality of (\ref{th3}) follows.\\
 The conditions
$$
T(v,v) \geq 0, \quad |I|^2 \geq0
$$
together form the {\it{dominant energy condition}}, which expresses that the energy density is non-negative and that the momentum $I$ is causal.
 The latter
condition physically means that the momentum $I$ propagates with a velocity
which is not greater than the velocity of light. \\
The so-called {\it{strongy energy condition}} also holds in the present situation, namely  in the form
$$
(m-2)T(v,v) \geq tr T.
$$
\begin{thm}\label{theor2}
 Consider the energy-momentum tensor $T= T_{ab}dx^adx^b$ to $\phi$ as above and lightlike vector fields $\l=\l^a\partial_a$, $n=n^a\partial_a$ such that $\l^an_a \equiv g_{ab}\l^an^b =1$. Then
\begin{equation}\label{R1}
 T(\l , \l ) \equiv T_{ab}\l^a \l^b = h(\phi_*\l , \phi_* \l ) \geq 0,
\end{equation}
and the one-form $I:= T(\cdot , \l)$ with components $I_a :=T_{ab}\l^b$ obeys
\begin{equation}\label{R2}
 0\leq |I|^2 \leq 2T(\l , \l )T(\l , n ).
\end{equation}
\end{thm}
{\bf{Proof}}:\\
Assertion (\ref{R1}) follows from
$$
T(\l , \l ) = (f_{ab} - eg_{ab})\l^a\l^b = f_{ab}\l^a\l^b = h_{ij}(\l^a\partial_a\phi^i)(\l^b\partial_b\phi^j).
$$
The projection tensor $p$ with components
$$
p_{ab}: = \l_an_b + n_a\l_b - g_{ab}
$$
is a useful tool. It is orthogonal to $\l$ and $n$, i.e.
$$
p_{ab}\l^b = p_{ab}n^b =0,
$$
and is positive semidefinite. Hence
$$
0 \leq p^{ab}I_aI_b = (2\l^an^b - g^{ab})I_aI_b = 2T(\l , \l )T(\l , n )- |I|^2,
$$
which proves the right-hand side inequality of (\ref{R2}). \\
Below we will also use
$$
p^{ab}T_{ab} = (2\l^an^b -g^{ab})T_{ab} = 2T(\l , n )- tr T = 2T(\l , n ) + (m-2)e.
$$
Let us, in order to complete the proof, consider a fixed point $x_0 \in X$ and use coordinates $x^a$ such that
$$
\l^a = \l^a_0, \quad n^a = \delta_1^a, \quad (p_{ab}) = diag (0, 0, 1, ..., 1)
$$
in that very point $x_0$, where $diag$ indicates a diagonal matrix. Formally the same $2\times 2$ subdeterminants of $(f_{ab})$ as in the preceding proof are non-negative. Their sum is now in another way transformed into a coordinate-invariant form
$$
p^{ab}[f_{ab}(\l^c\l^df_{cd}) - (\l^cf_{ac})(\l^d f_{bd})] \geq 0.
$$
Here we insert
$$
p^{ab}f_{ab} = p^{ab}(T_{ab} + eg_{ab}) = p^{ab}T_{ab} - (m-2)e = 2T(\l , n ),
$$
$$
\l^c\l^d f_{cd} = T(\l , \l ), \quad p^{ab} (\l^cf_{ac})(\l^df_{bd})=p^{ab}I_aI_b = 2T(\l , \l ) T(\l , n) - |I|^2.
$$
The result $|I|^2 = g^{ab}I_aI_b \geq 0$ follows.\\
Physically, theorem (4.2) can be interpreted in terms of a fictional observer
which moves faster and faster. In the limit, he reaches the velocity of light
and $v$ turns to $l$. The energy density $T(l,l)$ then remains non-negative and
the momentum $I$ remains causal.

\begin{cor}
 There holds $T(\l , n) \geq 0$. Especially,  $T(\l , n) =0$ iff $I_a = T(\l , \l )n_a$.
\end{cor}
{\bf{Proof}}:\\
The formulas (\ref{R1}), (\ref{R2}) imply $T(\l , n) \geq 0$. If $T(\l , n) =0$ then $|I|^2 = 0$ and
$$
0 = p^{ab}(\l^cf_{bc}) = p^{ab} I_b = (\l^a n^b + n^a \l^b - g^{ab})I_b = T(\l , \l )n^a - I^a
$$

\setcounter{equation}{0}
 \section{Implications of the Einstein equation}
Let us study
\begin{equation}\label{4.1}
 \kappa Ric =\phi^*h,
\end{equation}
for given background $(Y,h)$ as a relation between $g$ and $\phi$. Contraction with $g^{-1}$ yields
\begin{equation}\label{4.1-}
 \kappa R =2e.
\end{equation}
Double contraction  with a vector field $v=v^a\partial_a$ on $X$ yields
\begin{equation}\label{4.1--}
 \kappa Ric (v,v) = h(\phi_*v, \phi_*v).
\end{equation}
with the interpretation that $y = \phi(x)$ is to be inserted into the right-hand side of (\ref{4.1--});
$\phi_*v$ denotes the push-forward of  $v$  with respect to $\phi$.
As a conclusion, $\kappa Ric(v,v)$  is positive definite in $\phi_*v$ and is positive semi-definite in $v$.

\begin{prop}\label{pro4.1}
From (\ref{4.1}) there follows that $Ric$ and $d\phi$, interpreted as linear map, have the same rank:
\begin{equation}\label{4.2}
 r:= rank (Ric ) = rank (d\phi ).
\end{equation}
In particular:\\
 $r=0$ iff $\phi$ is constant.\\
 $r=m\equiv \dim X$ iff $\phi$ is an immersion.\\
 $r=n\equiv \dim Y$ iff $\phi$ is a submersion.\\
 $r=m=n$ iff $\phi$ is a local diffeomorphism.
\end{prop}
The {\it{proof}} of (\ref{4.2}) is an exercise in linear algebra.\\
 Note that $r=0$ means that $(X,g)$ is Ricci flat, i.e., $Ric =0$.
\begin{prop}\label{pro4.2}
 If  (\ref{4.1}) holds then the conditions
\begin{equation}\label{4.3}
 Ric (v,v)=0,
\end{equation}
\begin{equation}\label{4.4}
 Ric (\cdot , v)=0,
\end{equation}
\begin{equation}\label{4.5}
 \phi_*v=0,
\end{equation}
for a vector field $v=v^a\partial_a$ on $X$, are equivalent to each other. Moreover, they imply
\begin{equation}\label{4.6}
\pounds_v Ric =0,
\end{equation}
where $\pounds_v$ denotes the Lie derivative with respect to $v$.
\end{prop}
{\bf{Proof:}}\\
 Equation (\ref{4.1--})
in components reads
$$
\kappa R_{ab}v^av^b = h_{ij}(\phi_*v)^i(\phi_*v)^j,
$$
where $(\phi_*v)^i=v^a\partial_a\phi^i$. Moreover, (\ref{4.1}) implies
$$
\kappa R_{ab}v^b =h_{ij} (\partial_a\phi^i)(\phi_*v)^j.
$$
These formulas and the definiteness of $h$ give the first assertion. Next, we use comoving  coordinates $x^a$, which are  adapted to $v$, that means $v^a = \delta_0^a$, and we get
$$
(\phi^*v)^i = v^a\partial_a \phi^i = \partial_0\phi^i,
$$
$$
%\begin{array}{ccc}
 \kappa \pounds_vR_{ab} = \kappa\partial_0 R_{ab} = \partial_0(\partial_a\phi^i\partial_b\phi^jh_{ij}(\phi^k)).
%\end{array}
$$
If $\partial_0\phi^i=0$ then also $\partial_0R_{ab} =0$. This fact can be translated into the second assertion.
\begin{prop}\label{pro4.3}
 If the Ricci tensor vanishes on the vectors of some integrable distribution in the tangent bundle $TX$ and (\ref{4.1}) holds then $\phi$ is
constant on each leaf of the  foliation to the distribution.
\end{prop}
{\bf{Proof:}}\\
 A distribution of rank $s$ in $TX$ is integrable iff it admits adapted coordinates $x^a$, which means that the distribution is locally spanned by the coordinate vectors $\partial_1, \partial_2, ..., \partial_s$. The assumption becomes
$$
Ric (\partial_a, \partial_b)=0 \quad for\,\, a,b =1,2,..., s.
$$
Proposition (\ref{pro4.2}) then implies
$$
\partial_1\phi^i=0,\,\, \partial_2\phi^i =0,\,\, ..., \,\, \partial_s\phi^i=0.
$$
Hence  $\phi^i$  does not depend on $x^1, x^2, ..., x^s$ and
is constant if the point $x$  varies in a leaf of the foliation, i.e., if $x^{s+1}= const.,\,\, ...,\,\, x^m= const$.
%\end{proof}
\begin{prop}\label{pro4.5}
 The Einstein equation (\ref{4.1}) implies
\begin{equation}\label{4.7}
\kappa (\nabla_aR_{bc}+\nabla_bR_{ca} - \nabla_cR_{ab}) = 2h_{ij}(\nabla_{a}\partial_b\phi^i)(\partial_c\phi^j).
\end{equation}
\end{prop}
{\bf{Proof:}}\\
 Covariant differentiation of (\ref{4.1}) gives
$$
\kappa \nabla_cR_{ab} = h_{ij}[(\nabla_c\partial_a\phi^i)(\partial_b\phi^j)+(\partial_a\phi^i)(\nabla_c\partial_b\phi^j)].
$$
Some rearrangement yields  (\ref{4.7}).
%\end{proof}
\begin{prop}\label{pro4.6}
The Einstein equation (\ref{4.1}) implies
\begin{equation}\label{4.8}
h(tr(\nabla d\phi), d\phi)=0.
\end{equation}
This  equation  for a submersion $\phi$ implies  the harmonic or wave map equation (\ref{7--}).

\end{prop}
{\bf{Proof:}}\\
The Einstein tensor $Ric - \frac{1}{2}Rg$ is divergence-free. This fact and (\ref{4.7}) imply
$$
2h(tr(\nabla d\phi), d\phi)_c \equiv 2h_{ij} tr(\nabla d\phi)^i(\partial_c\phi^j)=\kappa g^{ab}(\nabla_a R_{bc}+\nabla_bR_{ca} - \nabla_cR_{ab})=0.
$$

If $\phi$ is a submersion, then the matrix with elements $h(\cdot , d\phi)_{ic}=h_{ij}\partial_c\phi^j$ has maximal rank and therefore (\ref{4.8}) implies (\ref{7--}).
%\end{proof}
\begin{prop}\label{pro4.7}
If the Ricci tensor is covariantly constant, i.e., $\nabla Ric = 0$, and the Einstein equation (\ref{4.1}) holds for a submersion $\phi$, then $\phi$ is totally geodesic, that means
\begin{equation}\label{12}
\nabla d\phi = 0.
\end{equation}
\end{prop}
{\bf{Proof:}}\\
 If $\nabla Ric =0$ then (\ref{4.7}) reduces to $h_{ij}(\nabla_a\partial_b\phi^i)(\partial_c\phi^j)=0$.
If, additionally, $\phi$ is a submersion, then (\ref{12}) follows by means of the rank argument already used in the preceding proof.\\

The next proposition needs some preparation. A diffeomorphism $f: X \longrightarrow X$ is called a {\it{Ricci symmetry}} iff
\begin{equation}\label{13}
f^*Ric =Ric.
\end{equation}
A vector field $v =v^a\partial_a$ on $X$ is called an {\it{infinitesimal Ricci symmetry}} or a {\it{Ricci collineation}} iff
\begin{equation}\label{14}
\pounds_vRic =0.
\end{equation}
It is well known that the flow $f_t$ for $t\in I$ of a Ricci collineation $v$ is a one-parameter family of Ricci symmetries, i.e. $f_t^* Ric = Ric$.
Here $x=f_t(x_0)$ by definition represents the solution of the initial-value problem
$$
\frac{dx}{dt} =v(x), \quad x|_{t=0}=x_0,
$$
and $I$ denotes an open interval which contains $0$.

\begin{prop}\label{pro4.8}
 Let $v=v^a\partial_a$ be a Ricci collineation of $(X, g)$. Then the Einstein equation (\ref{4.1}) implies that $\phi : X \longrightarrow Y$ is invariant under the flow $f_t$ for $t\in I$ of $v$, that means
\begin{equation}\label{15}
\phi \circ f_t =\phi.
\end{equation}
\end{prop}
{\bf{Proof:}}\\
 Let us again use comoving coordinates such that $v^a =\delta_0^a$.
 In these coordinates, $f_t$ is expressed by a translation
 $x^0 \mapsto x^0 + t$, $x^1 \mapsto x^1$, ..., $x^{m-1} \mapsto x^{m-1}$.
 We know already from the proof of Proposition 5.2  $\partial_0 \phi^i =0$, i.e. each $\phi^i$ is independent of $x^0$. Hence $\phi^i = \phi^i(x^a)$ does not change under $x^0 \mapsto x^0 + t$, which is just expressed by (\ref{15}).\\

The following definition is useful.\\
{\bf{Definition 5.1}} A property of subsets of a manifold $X$ holds {\it{globally}} if it is valid for every $X$. It holds {\it{locally}} if every point $x_0\in X$ has a neighborhood $U$ such that the property is valid for  $U$.
\begin{thm}\label{th4.2}
Let $\phi : X \longrightarrow Y$ be a submersion  between smooth manifolds $X$, $Y$ and let $Y$ be equipped with a positive definite metric $h$. \\
1-  Locally there is a positive definite metric $g$ on $X$ such that $\phi$ becomes a harmonic map  $(X, g)$ and $(Y, h)$.\\
2- Locally there is a Lorentzian metric $g$ on $X$ such that $\phi$ becomes a wave map.
\end{thm}
{\bf{Proof:}}\\
We consider (5.1) and use the fact that the problem of prescribed Ricci
curvature is locally solvable in the two cases \cite{19,20}. More precisely: cf.,
eg.\\
1- Einstein's equation (\ref{4.1}) locally has a positive definite solution $g$.
It can be constructed, e.g., through some boundary value problem \cite{19}. By assumption, $\phi$ is a submersion; Proposition \ref{pro4.6} implies $tr (\nabla d\phi ) =0$.\\
2- Einstein's equation (\ref{4.1}) locally has a Lorentzian solution $g$. It can be constructed,
 e.g., through some Cauchy initial value problem cf., e.g. \cite{20}.\\
  An argument like in item $1$ completes the proof.

\setcounter{equation}{0}
 \section{Product and warped product source manifolds}
{\bf{Definition 6.1}:} The {\it{product}} $(X, g)$ of two (pseudo-) Riemannian manifolds $({^{\prime}}X, {^{\prime}}g)$, $({^{\prime\prime}}X, {^{\prime\prime}}g)$ is given by $X= {^{\prime}}X \times\,{^{\prime\prime}}X$ as a product of manifolds and by
$$
g(u, v) = {^{\prime}}g({^{\prime}}u, {^{\prime}}v)\,+\, {^{\prime\prime}}g({^{\prime\prime}}u, {^{\prime\prime}}v)
$$
where $u, v$ are vector fields on $X$, ${^{\prime}}u, {^{\prime}}v$ are the push-forwards of $u, v$ with respect to the projection $X \longrightarrow {^{\prime}}X$, and ${^{\prime\prime}}u$, ${^{\prime\prime}}v$ are the push-forwards of $u, v$ with respect to the projection $X \longrightarrow {^{\prime\prime}}X$ \\
We write then
$$
g={^{\prime}}g \oplus \,{^{\prime\prime}}g,
$$
$$
 dim {^{\prime}}X = \, {^{\prime}}m, \quad dim {^{\prime\prime}}X = {^{\prime\prime}}m, \quad m={^{\prime}}m + \, {^{\prime\prime}}m.
$$
We apply the index convention
$$
{^{\prime}}a, \, {^{\prime}}b \, {^{\prime}}c, \,... = 1, 2, ...,{^{\prime}}m; \quad
{^{\prime\prime}}a, \, {^{\prime\prime}}b \, {^{\prime\prime}}c, \,... = {^{\prime}}m+1, \, {^{\prime}}m+2\, , ...,m.
$$
{\bf{definition 6.2}:} The {\it{warped product}} $(X, g)$ is given by $X= {^{\prime}}X \times\,{^{\prime\prime}}X$  like above and by
$$
g(u, v) = {^{\prime}}g({^{\prime}}u, {^{\prime}}v)\,+\, w^2\,\,{^{\prime\prime}}g({^{\prime\prime}}u, {^{\prime\prime}}v),
$$
where the warping function $w$ is a map  ${^{\prime}}X \longrightarrow \Re$ with positive values $w>0$.\\
 We write then
$$
g={^{\prime}}g \oplus \, w^2 \, {^{\prime\prime}}g,
$$

The following is known.
\begin{prop}
 If $(X, g)$ is the product of $({^{\prime}}X, \, {^{\prime}}g)$, $({^{\prime\prime}}X, \, {^{\prime\prime}}g)$ then the Einstein equation $\kappa Ric = \phi^* h \equiv h(d\phi , d\phi )$ decomposes into the two Einstein equations
\begin{equation}\label{7.1}
 \kappa {^{\prime}}Ric = h(\, {^{\prime}}d\phi, \, {^{\prime}}d\phi ), \quad \kappa {^{\prime\prime}}Ric = h(\, {^{\prime\prime}}d\phi, \, {^{\prime\prime}}d\phi )
\end{equation}
and the orthogonality condition with respect to $h$
\begin{equation}\label{7.2}
 h({^{\prime}}d\phi, \, {^{\prime\prime}}d\phi ) =0,
\end{equation}
in components
\begin{equation}\label{7.3}
\kappa \, {^{\prime}}R_{{^{\prime}}a\,{^{\prime}}b}=h_{ij} (\partial_{{^{\prime}}a}\phi^i)(\partial_{{^{\prime}}b}\phi^j),
\quad \kappa \, {^{\prime\prime}}R_{{^{\prime\prime}}a\,{^{\prime\prime}}b}=h_{ij} (\partial_{{^{\prime\prime}}a}\phi^i)(\partial_{{^{\prime\prime}}b}\phi^j),
\end{equation}
\begin{equation}\label{7.4}
h_{ij} (\partial_{{^{\prime}}a}\phi^i)(\partial_{{^{\prime\prime}}b}\phi^j) = 0.
\end{equation}
\end{prop}
\begin{prop}\label{pro7.2}
 If $(X, g)$ is the warped product of $({^{\prime}}X, \, {^{\prime}}g)$, $({^{\prime\prime}}X, \, {^{\prime\prime}}g)$ with warping function $w$ then the Einstein equation $\kappa Ric = \phi^* h \equiv h(d\phi , d\phi )$ decomposes into
\begin{equation}\label{7.5}
 \kappa (\,{^{\prime}}Ric - \, {^{\prime\prime}}m w^{-1}\, {^{\prime}}\nabla\,{^{\prime}}d w ) = h(\, {^{\prime}}d\phi, \, {^{\prime}}d\phi ),
\end{equation}
\begin{equation}\label{7.6}
 \kappa (\,{^{\prime\prime}}Ric - \, \frac{1}{{^{\prime\prime}}m} w^{2- {^{\prime\prime}}m }\, {^{\prime\prime}}\Delta w^{^{\prime\prime}m}\,{^{\prime\prime}}g)  = h(\, {^{\prime}}d\phi, \, {^{\prime\prime}}d\phi ),
\end{equation}
\begin{equation}\label{7.7}
 h({^{\prime}}d\phi, \, {^{\prime\prime}}d\phi ) =0.
\end{equation}
The trace parts of (\ref{7.5}), (\ref{7.6}) read
\begin{equation}\label{7.8}
 \kappa (\,{^{\prime}}R - \, {^{\prime\prime}}m w^{-1}\, {^{\prime}}\Delta w ) = 2\,\, {^{\prime}}e,
\end{equation}
\begin{equation}\label{7.9}
 \kappa (\,{^{\prime\prime}}R - \,  w^{2- {^{\prime\prime}}m }\, {^{\prime}}\Delta w^{^{\prime\prime}m}) = 2\,\, {^{\prime\prime}}e.
\end{equation}
\end{prop}
The proof is an exercise in higher differential geometry.
\par
We say that $\phi : {^{\prime}}X \times {^{\prime\prime}}X \longrightarrow Y$ does not depend on ${^{\prime}}x$ iff the restriction $\phi(\cdot , {^{\prime\prime}}x) : {^{\prime}}X \longrightarrow Y$ is a constant map for every ${^{\prime\prime}}x \in {^{\prime\prime}}X$. We say that $\phi$ does not depend on ${^{\prime\prime}}x$ with an analogous situation.
\begin{thm}\label{thm7.1}
 Let in the situation of proposition \ref{pro7.2} the restriction $\phi(\,{^{\prime}}x, \cdot) : {^{\prime\prime}}X \longrightarrow X$ be a submersion for every ${^{\prime}}x \in \, {^{\prime}}X$. Then $\phi$ does not depend on   ${^{\prime}}x$.
\end{thm}
{\bf{Proof}:} \\
If every $\phi(\,{^{\prime}}x, \cdot)$ is a submersion then the quantities $h_{ij}\partial_{{^{\prime\prime}}b}\phi^j$ in (\ref{7.4}) form a matrix of maximal rank, and (\ref{7.4}) implies $\partial_{{^{\prime}}a}\phi^i =0$, which gives the assertion.
\begin{thm}\label{thm7.2}
 Let in the situation of proposition \ref{pro7.2} the first factor $({^{\prime}}X, \, {^{\prime}}g)$ be properly Riemannian and closed. Then the following holds.\\
(i) The symmetric two-form $\kappa {^{\prime\prime}}Ric$ is positive semi-definite.\\
(ii) If $\kappa \,{^{\prime\prime}}Ric$ is not everywhere positive then $w=const.$\\
(iii) $\kappa \int_{{^{\prime}}X}w\, {^{\prime}}R d\, {^{\prime}}vol   \geq 0$.\\
(iv) If $\int_{{^{\prime}}X}w\, {^{\prime}}R d\, {^{\prime}}vol = 0$ then $w=const.$ and $\phi$ does not depend on ${^{\prime}}x$.
\end{thm}
{\bf{Proof}:}\\
 (i) Multiply (\ref{7.6}) by $w^{{^{\prime\prime}}m-2}$ and evaluate the two-forms at a vector field ${^{\prime\prime}}v \neq 0$ on ${^{\prime\prime}}X$:
\begin{equation}\label{S1}
\kappa ( w^{{^{\prime\prime}}m-2}\, {^{\prime\prime}}Ric ({^{\prime\prime}}v, {^{\prime\prime}}v )- \frac{1}{{^{\prime\prime}}m}({^{\prime}}\Delta w^{{^{\prime\prime}}m})|{^{\prime\prime}}v|^2)=w^{{^{\prime\prime}}m-2}h({^{\prime\prime}}d\phi ({^{\prime\prime}}v), {^{\prime\prime}}d\phi ({^{\prime\prime}}v)).
\end{equation}
Integrate this equation over ${^{\prime}}X$; the Laplacian term does not contribute, hence
$$
\kappa {^{\prime\prime}}Ric ({^{\prime\prime}}v, {^{\prime\prime}}v)\int_{{^{\prime}}X}w^{{^{\prime\prime}}m-2}d {^{\prime}}vol = \int_{{^{\prime}}X}w^{{^{\prime\prime}}m-2}h({^{\prime\prime}}d\phi ({^{\prime\prime}}v), {^{\prime\prime}}d\phi ({^{\prime\prime}}v))d{^{\prime}}vol.
$$
Both the integrals are non-negative, hence $\kappa {^{\prime\prime}}Ric ({^{\prime\prime}}v, {^{\prime\prime}}v) \geq 0$ for every ${^{\prime\prime}}v$.\\
(ii) If ${^{\prime\prime}}Ric ({^{\prime\prime}}v, {^{\prime\prime}}v)({^{\prime\prime}}x_0)= 0$ for some point ${^{\prime\prime}}x_0 \in {^{\prime\prime}}X$
 then
$$
\int_{{^{\prime}}X}w^{{^{\prime\prime}}m-2}h({^{\prime\prime}}d\phi ({^{\prime\prime}}v({^{\prime\prime}}x_0)), {^{\prime\prime}}d\phi ({^{\prime\prime}}v({^{\prime\prime}}x_0))d{^{\prime}}vol=0,
$$
which implies ${^{\prime\prime}}d\phi ({^{\prime\prime}}v({^{\prime\prime}}x_0)) = 0$. Evaluation of (6.10) at the point ${^{\prime\prime}}x_0$
reduces (\ref{S1}) to
$$
({^{\prime}}\Delta w^{{^{\prime\prime}}m})|{^{\prime\prime}}v({^{\prime\prime}}x_0)|^2= 0.
$$
We can assume
$|{^{\prime\prime}}v({^{\prime\prime}}x_0)|^2 \neq 0$; hence $w^{{^{\prime\prime}}m}$ is a harmonic function. A harmonic
function on a closed manifold is constant.\\
(iii) Multiply (6.8) by $w$ and integrate then over ${^{\prime}}X$. The Laplacian term
does not contribute; hence
$$
\kappa \int_{{^{\prime}}X}w\, {^{\prime}}R \, d\, {^{\prime}}vol=  \int_{{^{\prime}}X}v\, {^{\prime}}e\, d\, {^{\prime}}vol\geq 0.
$$
(iv) The last integral vanishes only if ${^{\prime}}e = 0$, which implies ${^{\prime}}d\phi = 0$, hence
$\phi(\cdot, {^{\prime\prime}}x)= const$ for fixed ${^{\prime\prime}}x \in  {^{\prime\prime}}X$. But then ${^{\prime\prime}}e(\cdot, {^{\prime\prime}}x) = const$ and the standard separation argument can be applied to
\begin{equation}\label{6.11}
\kappa w^{2-\, {^{\prime}}m}\, {^{\prime}}\Delta w^{{^{\prime\prime}}m}=\kappa \, {^{\prime\prime}}R- 2\, {^{\prime\prime}}e.
\end{equation}
Thus both sides of (6.11) equal a constant c. Integration of
$$
\kappa \, {^{\prime}}\Delta w^{{^{\prime\prime}}m}=cw^{{^{\prime\prime}}m-2}
$$
yields $c = 0$. Hence $w^{{^{\prime\prime}}m}$ is a harmonic function on the closed manifold ${^{\prime}}X$.
We find $w^{{^{\prime\prime}}m} =const$, $w =const$.

\begin{prop}\label{pro7.5}
 Let in the situation of the preceding theorem the second
factor $({^{\prime\prime}}X, \,{^{\prime\prime}}g)$ be properly Riemannian with vanishing scalar curvature:
${^{\prime\prime}}R = 0$. Then $w =const$ and $\phi$ does not depend on ${^{\prime\prime}}x \in {^{\prime\prime}}X$.
\end{prop}
{\bf{Proof:}}\\
Now equation (6.9) reduces to
$$
-\kappa \, {^{\prime}}\Delta w^{{^{\prime\prime}}m}=2w^{{^{\prime\prime}}m-2}\, {^{\prime\prime}}e
$$
and we have ${^{\prime\prime}}e\geq 0$. Integration over the closed manifold ${^{\prime}}X$ yields
$$
0=\int_{{^{\prime}}X}w^{{^{\prime\prime}}m-2}\, {^{\prime\prime}}e\,d\,{^{\prime}}vol,
$$
which implies
$$
{^{\prime\prime}}e={^{\prime\prime}}tr\, h({^{\prime\prime}}d\phi,\,{^{\prime\prime}}d\phi)=0.
$$
Hence ${^{\prime\prime}}d\phi = 0$, i.e. $\phi$ does not depend on ${^{\prime\prime}}x$, and $w^{{^{\prime\prime}}m}$ becomes a harmonic
function. We arrive at $w^{{^{\prime\prime}}m} =const$, $w = const$.\\

{\bf{Example 6.1.}} Let in the situation of proposition 6.2 the first factor
$({^{\prime}}X,\,{^{\prime}}g)$ be the unit circle $S^1$. It is a flat properly Riemannian closed manifold.
Theorem 6.4 implies that $w =const$ and $\phi$ does not depend on ${^{\prime}}x$.\\

{\bf{Example 6.2.}} A static metric $g = w^2dt^2-\,{^{\prime}}g$ can be interpreted as
a warped product metric on $X={^{\prime}}X\times{^{\prime\prime}}X$ where there the second factor
${^{\prime\prime}}X$ is one-dimensional. The proof of proposition 6.5 works, with a slight
modification, for this case. Hence Einstein's equation $\kappa Ric =\phi^*h$ implies
that $w =const$ and $\phi$ does not depend on $t\in {^{\prime\prime}}X$.

\setcounter{equation}{0}
 \section{Source manifolds with radiation conditions}

Lichnerowicz \cite{21} proposed conditions of pure radiation for a Lorentzian
manifold $(X,g)$:
$$
R_{abcd}l^d = 0, \quad  R_{ab[cd}l_{e]} = 0,
$$
where $R_{abcd}$ are the components of the Riemann curvature tensor and $l =
l^a\partial_a$ is a lightlike vector field. Bel \cite{22}  proposed weaker conditions:
$$
R_{abcd}l^bl^d = 0, \quad  l^bR_{ab[cd}l_{e]} = 0, \quad  l_{[f}R_{ab][cd}l_{e]} = 0.
$$
There are also radiation conditions for the Ricci tensor $Ric$ or for the Einstein
tensor $G :=Ric - \frac{1}{2}Rg$ with components
$$
G_{ab} :=R_{ab}- \frac{1}{2}Rg_{ab},
$$
namely
$$
R_{ab}l^b = 0, \quad  R_{a[b}l_{c]} = 0,
$$
$$
G_{ab}l^b = 0, \quad  G_{a[b}l_{c]} = 0.
$$
One of the present authors studied radiation conditions in \cite{23,24}.

\begin{thm}
 If the Einstein equation $\kappa Ric =\phi^*h$ holds then the conditions
\begin{equation}
G_{ab}l^b = 0,
\end{equation}
\begin{equation}
R_{a[b}l_{c]} = 0
\end{equation}
for a lightlike vector field $l =l^a \partial_a$ are equivalent to each other.
\end{thm}
{\bf{Proof:}}\\
There is another lightlike vector field $n = n^a \partial_a$ such that $l^an_a = 1$. Then
the sum $v := l+n$ is timelike with $v^av_a = 2$ and $g^+_{ab} = v_av_b-g_{ab}$ are the
components of a positive definite metric $g_+$.\\
Let us consider
$$
\psi^i_a := 2v^bl_{[a}\partial_{b]}\phi^i \equiv l_a(\phi_*v)^i -\partial_{a}\phi^i
$$
and calculate
$$ g^{ab}_+ h_{ij}\psi^i_a \psi^j_b =-g^{ab}h_{ij}\psi^i_a\psi^j_b =h_{ij}(\phi_*v)^i(\phi_*l)^j-e =\kappa G_{ab}v^al^b.
$$
Here we made use of
$$
g^{ab}_+ =v^av^b-g^{ab}, \quad  v^a\psi^i_a = 0
$$
and of $\kappa Ric=\phi_*h$. If $G_{ab}l^b = 0$ then the above positive definite expression
vanishes and we get
$$
\psi^i_a = 0, \quad \partial_a\phi^i =l_a(\phi_*v)^i,
$$
$$
\kappa R_{ab}=h_{ij}(\partial_a\phi^i)(\partial_{b}\phi^j)=l_al_bh(\phi_*v,\phi_*v).
$$
Conversely, from $R_{ab} = rl_al_b$ with some scalar $r$ there follow $R = 0$ and
(7.1).\\

We here dwell again on the fact that  to every smooth vector field
$v =v^a\partial_a$ on $X$ there are comoving coordinates $x^a = x^0, x^1, ..., x^{m-1}$,
which means $v^a = \delta^a_0$. The adaptedness of coordinates is
preserved by coordinate transformations of the form
\begin{equation}\label{6.1}
 \bar{x}^0 = x^0 + f(x^i), \quad \bar{x}^i = \bar{x}^i(x^j).
\end{equation}

We apply in this section  the index convention
$$
a, b, c, ... = 0, 1, 2, ..., m-1,
$$
$$
i, j, k, ... = 1, 2, ..., m-1,
$$
$$
I, J, K, ... = 2, 3, ..., m-1.
$$
\begin{prop}\label{pro6.1}
 A Lorentzian manifold $(X, g)$ admits a lightlike\\
...  hypersurface-orthogonal Killing vector field $\l = \l^a\partial_a$, that means
$$
\l_{[c}\nabla_a\l_{b]}=0, \quad \nabla_{(a}\l_{b)} =0,
$$
 iff in coordinates adapted to $\l$ the metric assumes the form
\begin{equation}\label{6.2}
g=2g_{01}dx^0dx^1  + g_{ij}dx^idx^j,
\end{equation}
where $g_{01}$, $g_{ij}$ do not depend on $x^0$.
The component $g_{01} = g_{01}(x^k)$ is invariant under gauge transformations
(\ref{6.1}) and the part $g_{IJ}dx^Idx^J$ of (\ref{6.2}) shows tensorial behavior
under the part $\bar{x}^i = \bar{x}^i(x^j)$ of (\ref{6.1}).
Moreover, the matrix $(g_{IJ}) = (g_{IJ}(x^k))$ is  negative definite.\\
... covariantly constant vector field $\l = \l^a\partial_a$, that means
$\nabla_a\l_b =0$, iff in coordinates adapted to $\l$ the metric assumes the form
\begin{equation}\label{6.3}
g=2dx^0dx^1  + g_{ij}dx^idx^j,
\end{equation}
where the components $g_{ij}$ do not depend on $x^0$.\\
... covariantly constant vector field $\l = \l^a\partial_a$ such that the Bel condition
\begin{equation}\label{6.4}
\l_{[e}R_{ab][cd}\l_{f]}=0
\end{equation}
holds iff there are coordinates adapted to $\l$ such that
\begin{equation}\label{6.5}
g=2g_{01}dx^0dx^1  + g_{11}(dx^1)^2 + 2g_{1I}dx^1dx^I - dx^Idx^I,
\end{equation}
where $g_{11}$, $g_{1I}$ do not depend on $x^0$ and summation over $I$ is applied.\\
... covariantly constant vector field $\l = \l^a\partial_a$ such that the Lichnerowicz condition
\begin{equation}\label{6.6}
\l_{[e}R_{ab]cd}=0
\end{equation}
holds iff there are coordinates adapted to $\l$ such that
\begin{equation}\label{6.7}
g=2g_{01}dx^0dx^1  + g_{11}(dx^1)^2  - dx^Idx^I,
\end{equation}
where $g_{11}$ does not depend on $x^0$ and summation over $I$ is applied.
\end{prop}
All these facts together with {\it{proof}}  and additional information are given in the papers \cite{23,24}.
\par
A Lorentzian manifold $(X, g)$ which admits a covariantly constant lightlike vector
$\l = \l^a\partial_a$ is called a {\it{plane-fronted gravitational wave with parallel rays}},
abbreviated {\it{pp-wave}}. Note that from $\nabla_a\l_b =0$ and the Ricci identity there follows the Lichnerowicz condition
$$
R_{abcd}\l^d =0.
$$
\begin{thm}
 A metric (\ref{6.2}) satisfies an Einstein equation $\kappa Ric = \phi^*h$ iff $g_{01}=g_{01}(x^1, x^K)$ is a harmonic function of $x^K=x^2, x^3, ..., x^{m-1}$ with respect to the positive definite metric (which depends on $x^1$ as a parameter) $-g_{IJ}dx^Idx^J$.
\end{thm}
{\bf{Proof}:}\\
 Some calculation gives the components
$$
R_{00}=0, \quad R_{01} = \frac{1}{2}\Delta g_{01}
$$
of the Ricci tensor $Ric = R_{ab}dx^adx^b$, where $\Delta$ denotes the Laplace operator with respect to $-g_{IJ}dx^Idx^J$.
By proposition 5.2, from $R_{00} \equiv R_{ab}\l^a\l^b =0$ and $\kappa Ric = \phi^*h$ there follows $R_{01} \equiv R_{1b}\l^b =0$. The assertion follows.
\begin{thm}
 Let $m=dim X =4$. A metric of the form  (\ref{6.5}) satisfies an Einstein equation $\kappa Ric = \phi^*h$ iff it satisfies (\ref{6.6}), that means iff it can be brought into the form  (\ref{6.7}).
\end{thm}
{\bf{Proof}:}\\
 Calculation of Ricci components gives $R_{IJ} = 0$; in particular $R_{II} =0$ (without summation). By Proposition 5.2, from this and $\kappa Ric = \phi^*h$ there follows $R_{1I} =0$. For $m = 4$, there are only two independent curvature components of type $R_{iJKL}$, namely
$$
R_{1223}= - R_{13}, \quad R_{1323} = R_{12}.
$$
Thus we get $R_{iJK2}= 0$ which is expressed by (\ref{6.7}) in a coordinate invariant way.

\section{Discussion}
The literature on harmonic or wave maps is very extensive. There are good
surveys on harmonic maps \cite{4,5}. Work on such maps in the role of matter
fields coupled with gravitation began about 1980 \cite{10,11}. One of the authors
of this paper worked, with coauthors, already on this subject; we refer to
the paper \cite{17} and the unpublished preprint \cite{25}.
\par
The Einstein equation $\kappa Ric = \phi^*h$ or $\kappa G = T$, where $G = Ric -\frac{1}{2}Rg$
denotes the Einstein tensor and $T = \phi^*h-eg$ the energy-momentum tensor
of $\phi$ exhibits some remarkable properties:\\
- The rank of $Ric$, taken as a linear map, equals the rank of the differential
$d\phi$ (Proposition 5.1).\\
- The symmetries of the Ricci tensor $ Ric$ and of the map $\phi$ are closely related
to each other (Propositions 5.2, 5.3, 5.7).\\
- If $\phi$ is submersive then the Einstein equation implies the harmonic or wave
map equation (Proposition 5.5; cf. also Proposition 5.6).\\
- In the Lorentzian case there are identities and estimates for the energy momentum tensor $T$
which indicate a physically good behavior of $T$ (Proposition
3.2, Theorems 4.1, 4.2).\\
- In the Lorentzian case there is a tendency to enhance radiation conditions.\\
That means, the Einstein equation leads from one condition to a stronger
condition (Section 7).
\par
There is also a situation where the Einstein equation serves as an auxiliary
construction: for a given submersion $\phi$ there locally exists a metric $g$
on $X$ which makes $\phi$ to a harmonic or wave map (Theorem 5.8).\\
One paper cannot touch all aspects of a subject. We did not discuss here:\\
- Bochner-Weitzenb\"{o}ck technique \cite{5,13,14,15},\\
- consequences of second variation formulas,\\
- factorizations of the map $\phi$ \cite{18},\\
- exact solutions \cite{7,8,9,10,11,12,13,14,15,16},\\
- coupling of $\phi$ to a gravitational theory other than Einstein's theory.\\

These topics are by far not exhausted and could be subjects of further
research.

\section*{Acknowledgments}
Former discussions and cooperations with T. Hirschmann and T. Deck are
gratefully acknowledged. We thank the professors G. Huisken and A. D.
Rendall for inspiring discussions.

%\begin{thebibliography}{widest-label}

\end{document}